\documentclass{wiley-article}
\usepackage{siunitx}
\usepackage{booktabs}
\usepackage[flushleft]{threeparttable}
\usepackage{multirow}
\usepackage{algorithm}
\usepackage{algpseudocode}
\usepackage{tabularx}
\usepackage[titletoc,toc,title]{appendix}

\usepackage{caption}

\usepackage[numbers]{natbib}
\title{A Novel Two-stage Deming Regression Framework with Applications to Association Analysis between Clinical Risks}

\author[1]{Yajie Duan}
\author[1]{Javier Cabrera}
\author[2]{Davit Sargsyan}

\affil[1]{Department of Statistics, Rutgers, The State University of New Jersey, Piscataway, NJ, USA}
\affil[2]{Ernest Mario School of Pharmacy, Rutgers, The State University of New Jersey, Piscataway, NJ, USA}

\corraddress{Javier Cabrera}
\corremail{cabrera@stat.rutgers.edu}

\begin{document}

\begin{frontmatter}
\maketitle

\begin{abstract}
In healthcare, clinical risks are crucial for treatment decisions, yet the analysis of their associations is often overlooked. This gap is particularly significant when balancing risks that are weighed against each other, as in the case of atrial fibrillation (AF) patients facing stroke and bleeding risks with anticoagulant medication. While traditional regression models are ill-suited for this task due to standard errors in risk estimation, a novel two-stage Deming regression framework is proposed to address this issue, offering a more accurate tool for analyzing associations between variables observed with errors of known or estimated variances. The first stage is to obtain the variable values with variances of errors either by estimation or observation, followed by the second stage that fits a Deming regression model potentially subject to a transformation. The second stage accounts for the uncertainties associated with both independent and response variables, including known or estimated variances and additional unknown variances from the model. The complexity arising from different scenarios of uncertainty is handled by existing and advanced variations of Deming regression models. An important practical application is to support personalized treatment recommendations based on clinical risk associations that were identified by the proposed framework. The model's effectiveness is demonstrated by applying it to a real-world dataset of AF-diagnosed patients to explore the relationship between stroke and bleeding risks, providing crucial guidance for making informed decisions regarding anticoagulant medication. Furthermore, the model's versatility in addressing data containing multiple sources of uncertainty such as privacy-protected data suggests promising avenues for future research in regression analysis.

\end{abstract}
\end{frontmatter}

\section{Introduction}
The assessment of clinical risks is a fundamental aspect of medication and treatment. While clinical risk prediction tools are widely used, there has been a lack of focus on analyzing the connection between these risks, which are crucial for making informed decisions about medication. Especially when dealing with clinical outcomes that must be balanced, such as disease severity and drug side effects, understanding the associations between their risks becomes essential for treatment choices \cite{dimarco2005factors}. Based on the relationship between clinical risks, precision medicine recommendations could be provided for aligning individual patient risks with the underlying association \cite{goetz2018personalized, sargsyan2022assessment}. For instance, patients with atrial fibrillation (AF) have an increased risk of stroke. The AF may form blood clots in the heart atrium, which would travel to the brain and cause stroke \cite{ivuanescu2021stroke, sharma2015efficacy}. Typically, these patients are treated with anticoagulants to inhibit the formation of blood clots. However, it's essential to note that anticoagulants may elevate the risk of major bleeding \cite{pengo2001oral, dimarco2005factors}. Thus, when considering the treatment for an individual patient, a careful evaluation is necessary to weigh the risk of stroke against that of bleeding. Analyzing the association between these two risks is crucial in making informed decisions about prescribing anticoagulants. There are existing risk scoring systems developed independently, such as $\text{\emph{CHADS}}_2$\cite{gage2001validation} and its extension $\text{\emph{CHA}}_2\text{\emph{DS}}_2\text{\emph{-VAS}}_c$ \cite{lip2010refining} to help predict the one-year risk of stroke after the initial diagnosis of atrial fibrillation, and $\text{\emph{HEMORR}}_2\text{\emph{HAGES}}$ and \emph{HAS-BLED} \cite{pisters2010novel} to assess bleeding risks in patients on oral anticoagulants. However, no toWLS have specifically addressed the relationship between these two risks, which is essential for decision-making about anticoagulation \cite{sargsyan2022assessment}.

The analysis of associations between clinical risks is crucial for making informed medication decisions, yet there has been limited attention to developing a comprehensive tool for this purpose. To explore the relationships between multiple risks, the regression models are valuable, involving the prediction of risks based on covariates such as patient demographics and clinical data, followed by fitting a regression line to the estimated risks. However, it's important to note that these risks are estimated using statistical models, resulting in values that are estimates with associated standard errors rather than observed data \cite{efron1981nonparametric}. Consequently, traditional regression models, which assume fixed values for covariates \cite{draper1998applied, kiers1997weighted}, are not suitable for this scenario where the risks serve as both covariates and responses and are estimated with associated standard errors.

This issue, initially motivated by the association analysis of clinical risks, has broader implications and can be extended to address regression between variables that are estimated with standard errors, or observed with known standard deviations or with other known variability. The standard errors may come from a statistical model estimating the variable, and the other types of measurement may come with known standard deviations, or with error terms with known variances such as data with privacy errors. For example, creatine and LDL from blood tests are estimates with standard errors. In the case of patients with AF, the risks of stroke and bleeding are the variables for which the values are to be estimated with corresponding variances of prediction errors, i.e., the squared standard error of estimates. In this paper, the known or estimated variances of errors in the variables will be used to represent both the squared known standard deviations of error terms such as privacy errors, and the squared standard errors of the estimates of variables. 

To address the regression between variables that are observed with errors of known or estimated variances, a novel two-stage Deming regression model was introduced in this work. Deming regression, originally proposed by Deming in 1943 \cite{deming1943statistical}, is an error-in-variables model used to determine the best-fit line for variables, considering errors in both the $\mathrm{x}$ - and $\mathrm{y}$-axes. This differs from conventional regression models that only consider errors in the response variable. Deming regression is commonly employed in clinical chemistry for method comparison studies, aiming to identify systematic differences between two measurement methods \cite{linnet1998performance,martin2000general, saraccli2013comparison}. In its standard form, Deming regression assumes constant error variations among observations and estimates regression parameters using the least-squares approach with a predefined ratio between error variances of dependent and independent variables \cite{cornbleet1979incorrect}. Building upon this foundation, Linnet \cite{linnet1990estimation} proposed a weighted modification of the estimation procedure, which accounts for the proportional relationship between error variances and actual variable levels. In cases where the uncertainties of each observation are known, Deming \cite{deming1943statistical} formulated the generalized Deming regression model, further developed by York \cite{york1966least} to provide an exact analytic solution. Williamson \cite{martin2000general} later refined York's work, deriving equations for the linear regression parameters and their associated standard errors.

When dealing with the regression between variables observed with errors of known or estimated variances, such as the association analysis between clinical risks, it is essential to carefully account for uncertainties in both the independent and response variables across various scenarios. The known or estimated variances may or may not be enough to explain the overall uncertainty of the regression model, and additional error terms may be needed. To address this challenge, a novel two-stage Deming regression framework is proposed in this paper, which is based on the existing errors-in-variables models. In section 2, the standard Deming models are presented first, followed by the methodology of the new two-stage Deming framework. To illustrate the applications of this proposed methodology, in section 3 we analyzed a case study of the relationship between stroke and bleeding risks among patients with atrial fibrillation. In section 4, a discussion on future research directions is provided.

\section{Methodology}
\subsection{Deming regression}
Deming regression is a technique employed to establish a linear relationship within two-dimensional data, where both $\mathrm{X}$ and $\mathrm{Y}$ variables are subject to measurement errors \cite{linnet1990estimation}. The paired observations, $\left(x_{i}, y_{i}\right)$, measured with errors, $\varepsilon_{i}$ and $\delta_{i}$, are used to fit the regression line between the true (expected) values $X_{i}$ and $Y_{i}$. The regression equation between true values is
\begin{equation}
Y_{i}=\beta_{0}+\beta_{1} X_{i}
\end{equation}
The intercept, $\beta_{0}$, and the slope, $\beta_{1}$, are estimated based on the observed values $x_{i}$ and $y_{i}$ where
$$
\begin{aligned}
x_{i} & =X_{i}+\varepsilon_{i} \\
y_{i} & =Y_{i}+\delta_{i} .
\end{aligned}
$$
It is assumed that the error terms $\varepsilon_{i}$ and $\delta_{i}$ are independent normal variables with expected values of zero \cite{linnet1990estimation}. In a simple Deming regression model, the error variances are assumed to be constant among observations. However, if the uncertainties vary and are known for each observation, a generalized Deming regression model can be employed to estimate the regression parameters \cite{martin2000general}.

For the simple Deming regression model, it is assumed constant error variances for each variable among the observations. Let $\lambda$ be the constant ratio of the two error variances such that $\lambda=\frac{\operatorname{Var}(\varepsilon)}{\operatorname{Var}(\delta)}$. The least-squares estimation of the Deming regression model minimizes the sum of squares \cite{linnet1990estimation}
\begin{equation}
\label{eq:eq2}
S S=\sum_{i=1}^{N}\left[\left(x_{i}-X_{i}\right)^{2}+\lambda\left(y_{i}-Y_{i}\right)^{2}\right]
\end{equation}
The slope estimate, $\hat{\beta}_{1}$, and the intercept estimate, $\hat{\beta}_{0}$, are computed as 
\begin{equation}
\label{eq:eq3}
\hat{\beta}_{1}=\frac{(\lambda q-u)+\sqrt{(u-\lambda q)^{2}+4 \lambda p^{2}}}{2 \lambda p}, \hat{\beta}_{0}=\bar{y}-\hat{\beta}_{1} \bar{x}
\end{equation}
where $u=\sum_{i=1}^{N}\left(x_{i}-\bar{x}\right)^{2}, q=\sum_{i=1}^{N}\left(y_{i}-\bar{y}\right)^{2}, p=\sum_{i=1}^{N}\left(x_{i}-\bar{x}\right)\left(y_{i}-\bar{y}\right)$ \cite{linnet1990estimation}. The true values are estimated from the regression coefficients and $\lambda$ as $\hat{X}_{i}=x_{i}+\frac{\lambda \hat{\beta}_{1} d_{i}}{\left(1+\lambda \hat{\beta}_{1}^{2}\right)}, \hat{Y}_{i}=y_{i}-\frac{d_{i}}{\left(1+\lambda \hat{\beta}_{1}^{2}\right)}$ with $d_{i}=y_{i}-\left(\hat{\beta}_{0}+\hat{\beta}_{1} x_{i}\right)$ \cite{jensen2007deming}.

For the generalized Deming model, it is assumed that the uncertainties for each observation are known and denoted by $u_{i}=\operatorname{Var}\left(\varepsilon_{i}\right), v_{i}=$ $\operatorname{Var}\left(\delta_{i}\right)$. Then the least-squares approach minimizes the sum of squares \cite{deming1943statistical}
\begin{equation}
\label{eq:eq4}
\chi^{2}=\sum_{i=1}^{N}\left[\frac{\left(x_{i}-X_{i}\right)^{2}}{u_{i}}+\frac{\left(y_{i}-Y_{i}\right)^{2}}{v_{i}}\right]
\end{equation}
Building upon Deming's fundamental concepts, York \cite{york1966least} laid the groundwork for a comprehensive and precise approach to address the problem of estimating the regression parameters for the generalized Deming model. Williamson \cite{martin2000general} subsequently rectified and enhanced York's contributions, resulting in the derivation of equations for the linear regression parameters and their SEs. As presented by Williamson \cite{martin2000general} , the slope and intercept are given by:
\begin{equation}
\label{eq:eq5}
\hat{\beta}_{1}=\frac{\sum_{i=1}^{n} w_{i} z_{i} y_{i}^{\prime}}{\sum_{i=1}^{n} w_{i} z_{i} x_{i}^{\prime}}, \hat{\beta}_{0}=\bar{y}_{w}-\hat{\beta}_{1} \bar{x}_{w}
\end{equation}
where $w_{i}=\left[v_{i}+\left(\hat{\beta}_{1}\right)^{2} u_{i}\right]^{-1}, \bar{x}_{w}=\frac{\sum_{i=1}^{n} w_{i} x_{i}}{\sum_{i=1}^{n} w_{i}}, \bar{y}_{w}=\frac{\sum_{i=1}^{n} w_{i} y_{i}}{\sum_{i=1}^{n} w_{i}}, x_{i}^{\prime}=x_{i}-\bar{x}_{w}, y_{i}^{\prime}=y_{i}-\bar{y}_{w}, z_{i}=w_{i}\left(v_{i} x_{i}^{\prime}+\hat{\beta}_{1} u_{i} y_{i}^{\prime}\right)$. An iterative calculation procedure is required to obtain the estimates.

\subsection{Two-stage Deming regression framework}
\subsubsection{Model framework}

For regression between variables that are observed with errors of known or estimated variances, e.g., observed values with the squared known standard deviations or estimates with the squared standard errors, the underlying uncertainties of the true (expected) values $X_{i}$ and $Y_{i}$ for each observation may or may not be adequately accounted for by the known or estimated variances. In this case, the Deming model needs to be extended to introduce an extra error term to explain the additional unknown variability. Moreover, for variables that require estimation or prediction, the values $x_{i}$ and $y_{i}$ correspond to the estimated values from a predictive model with their associated variances of prediction errors for each observation, i.e., the squared standard errors of estimates. 

Putting all these together, the initial process of obtaining the values with variances of errors in the variables either by estimation or observation, followed by the Deming model considering the known and unknown error terms, results in a new framework called Two-stage Deming regression. 

The first stage of this new framework produces estimated or observed values of $x_{i}$ and $y_{i}$ together with their corresponding variances of errors in the variables. These values of $x_{i}$ and $y_{i}$ might be estimated from a predictive model using existing data, or be directly observed with known standard deviations or with errors of known variances such as data with privacy errors. The squared standard errors of the estimates, or the squared known standard deviations were plugged in as if they were known or estimated variances of the errors associated with variables. At the end of the first stage, the observations are paired with their errors of known or estimated variances. Following this, in the second stage, the objective is to estimate the relationship between the variables with errors from the first stage, potentially subject to a transformation, using a Deming model.

The framework of this model is outlined in Algorithm 1. In this work, we focus on the univariate Deming regression case.
\begin{algorithm}[H]
\caption{Two-stage Deming regression Framework (Univariate)}
\begin{algorithmic}
\State \textbf{{1st Stage: }} Obtain observed or estimated values of variables $z_i, w_i$ with errors $e_{zi}, e_{wi}$ and their variances $\sigma^2_{zi}, \sigma^2_{wi}$ for observations $i = 1,..., n$.
\State \textbf{{2nd Stage: }} Fit Deming regression model.
\State \quad  \textbf{Step 1: } To obtain a linear relationship, take transformations $x_i = f_1(z_i), y_i = f_2(w_i)$ with errors $e_{xi}, e_{yi}$ and their variances $\sigma^2_{xi}, \sigma^2_{yi}$.
\State \quad  \textbf{Step 2: } Consider ${Y}_{i}=\beta_{0}+\beta_{1}{X}_{i},$ and three scenarios:
\State \quad \quad \quad \textbf{A.} $x_{i}=X_{i}+\varepsilon_{i}, y_{i}=Y_{i}+\delta_{i} \longrightarrow$ fit a simple Deming regression model
\State \quad \quad \quad \textbf{B.} $x_{i}=X_{i}+e_{xi}, y_{i}=Y_{i}+e_{yi} \longrightarrow$ fit a generalized Deming with LSM
\State \quad \quad \quad \textbf{C.} $x_{i}=X_{i}+e_{xi}+\varepsilon_{i}, y_{i}=Y_{i}+e_{yi}+\delta_{i} \longrightarrow$ fit a generalized Deming with MLE
\end{algorithmic}
\end{algorithm}

Suppose the relationship between variables $z$ and $w$ is of interest. To establish the relationship, the new framework displayed in Algorithm 1 consists of two distinct stages. In the first stage, the values of $z$ and $w$ are either observed with known variances $\sigma^2_{z}$ and $\sigma^2_{w}$ of the error terms $e_{z}$ and $e_{w}$ such as measurement errors or privacy errors, or estimated by models. In the latter case, prediction models are fitted based on independent covariates to estimate the concerned variables, $z$ and $w$, with associated prediction errors, denoted as $e_{z}$ and $e_{w}$, respectively. It's worth noting that the variables $z$ and $w$ can be predicted independently using different covariates or jointly using the same independent variables. In either case, it is imperative to ensure that the concerned variables $z$ and $w$ are paired and associated with the same subject, for the purpose of exploring their relationship. The variances of the prediction errors, represented as $\sigma_{z}^2$ and $\sigma_{w}^2$, are the squared standard errors of estimates, and can be estimated for each observation by the prediction models. Subsequently, in the second stage, a Deming regression model is considered under various situations, taking into account the obtained values along with errors of known or estimated variances.

In the second stage, two essential steps are involved in fitting the model. First, to establish a linear Deming regression line, the initial step is to apply appropriate transformations, such as log transformation or other non-linear monotonic functions, to the observed or predicted values of the concerned variables, $z$ and $w$, to achieve a linear relationship. These transformed variables are denoted as $x=f_1(z)$ and $y=f_2(w)$. Following this, the variances of the error terms $e_{x i}, e_{y i}$ for the transformed variables, can be approximated using the delta method \cite{oehlert1992note, doob1935limiting} based on the variances $\sigma_{z i}^{2}$ and $\sigma_{w i}^{2}$ obtained in the first stage. 

Once the transformed values and their error term variances are determined, the second step entails fitting a Deming regression model, considering the linear relationship between these transformed variables, i.e., $Y_{i}=\beta_{0}+$ $\beta_{1} X_{i}$. It's important to note that the disparities between the true values $\left(X_{i}, Y_{i}\right)$ and the transformed values $\left(x_{i}, y_{i}\right)$ may or may not be adequately explained by the errors from the first stage. Therefore, in the second step of fitting a Deming regression line between the true values $\left(X_{i}, Y_{i}\right)$, three distinct scenarios listed in Algorithm 1 were considered for handling the error terms:
\begin{itemize}
\item \textbf{Scenario A} considers the presence of unknown measurement errors $\varepsilon_{i}$ and $\delta_{i}$ with constant variances among observations, while assuming the errors associated with each observation from the initial stage are negligible.
\item \textbf{Scenario B} takes into account only the uncertainties of each observation derived from the first stage, denoted as $e_{x i}$ and $e_{y i}$, for the model fitting.
\item \textbf{Scenario C} incorporates both sets of error terms originating from the first stage and additional unknown variances into the model fitting. It's a combination of Scenario A and B.
\end{itemize}
The three scenarios tackle the uncertainties surrounding transformed values in distinct ways. The adequacy of the known or estimated variances from the first stage to account for the inherent disparities between true values $\left(X_{i}, Y_{i}\right)$ and observed values $\left(x_{i}, y_{i}\right)$ depends on the variances of the error terms $e_{x i}$ and $e_{y i}$. If these variances are sufficient to account for the uncertainties, they could be directly utilized to fit a Deming regression line between true values, as illustrated in Scenario B. However, if the known or estimated variances are an order of magnitude smaller than the underlying uncertainties, they may be disregarded during the model fitting process, and an assumption of constant unknown variances among observations is made, as Scenario A does. In cases where the known or estimated variances are insufficient to explain the underlying uncertainties but comparable in order of magnitude, neglecting them during the model fitting would result in a loss of information and variability for each observation, and the results may be biased. In such instances, Scenario C combines both errors with known and unknown variances to account for the overall uncertainties.

To evaluate and compare the known or estimated variances with the underlying overall uncertainties, a formal approach involves conducting a likelihood ratio test (LRT) \cite{buse1982likelihood} among models fitted under the three scenarios. However, when dealing with large datasets, Scenario C would be typically favored by the test. A practical method for selecting the most suitable scenario entails initially fitting the model under Scenario B, which exclusively accounts for uncertainties stemming from the known or estimated variances, and then conducting an evaluation of residuals from the fitted line. This evaluation relies on a criterion based on the ratio between the mean value of square roots of the known or estimated variances, and residual standard deviation from the fitted model under Scenario B, which is defined as
\begin{equation}
\label{eq:eq6}
r=\frac{\overline{\sqrt{\sigma^2_{yi}}}}{\sigma_{e}}
\end{equation}
where $\sigma^2_{y i}$ signifies the known or estimated variances of $y_i$ for each observation from the first stage, and $\sigma_{e}$ represents the residual standard deviation from the fitted Deming regression model under Scenario B. The selection of scenarios could be guided by the value of $r$: if $r$ closely approaches 0, indicating that the known or estimated variances are an order of magnitude smaller than the underlying uncertainties, Scenario A is the preferred choice, disregarding the error terms with known or estimated variances. Conversely, if $r$ approaches or exceeds 1, Scenario B is the optimal selection as the known or estimated variances account for the uncertainties well in the model. Otherwise, Scenario $\mathrm{C}$ is chosen to include both error terms with known and unknown variances.

\subsubsection{Model fitting}

In the second stage of the proposed framework in Algorithm 1, to establish the regression line under various scenarios, different errors-in-variables models are employed to estimate the regression parameters. 

In Scenario A, unknown measurement errors $\varepsilon_{i}$ and $\delta_{i}$ are considered with constant variances disregarding the errors with known or estimated variances for each observation from the first stage, i.e.,
$$
\begin{aligned}
x_{i} & =X_{i}+\varepsilon_{i} \\
y_{i} & =Y_{i}+\delta_{i} .
\end{aligned}
$$
To estimate the linear relationship between true values $\left(X_{i}, Y_{i}\right)$, i.e., $Y_{i}=\beta_{0}+\beta_{1} X_{i}$, the simple Deming regression model could be fitted with a pre-determined constant ratio $\lambda$ between the variances of two errors $\varepsilon_{i}$ and $\delta_{i}$. This model is fitted by the least squared method, as indicated in Equations \eqref{eq:eq2} and \eqref{eq:eq3}. 

In Scenario B, where we only consider the errors $e_{x i}$ and $e_{y i}$ with known or estimated variances for each observation from the first stage, denoted as:
$$
\begin{aligned}
& x_{i}=X_{i}+e_{x i} \\
& y_{i}=Y_{i}+e_{y i}
\end{aligned}
$$
the generalized Deming regression model would help establish the regression line through the least squared method with an iteration calculation procedure, as depicted in Equations \eqref{eq:eq4} and \eqref{eq:eq5}. 

For Scenario C, it is characterized by the combination of known and unknown errors, i.e.,
$$
\begin{aligned}
& x_{i}=X_{i}+e_{x i}+\varepsilon_{i} \\
& y_{i}=Y_{i}+e_{y i}+\delta_{i}
\end{aligned}
$$
where $\left(x_{i}, y_{i}\right)$ are the transformed values with errors $\left(e_{x i}, e_{y i}\right)$ and their known or estimated variances, $\left(X_{i}, Y_{i}\right)$ are the true (expected) values, and $\left(\varepsilon_{i}, \delta_{i}\right)$ are the additional errors terms to account for the overall uncertainties. The regression parameters can be estimated by a generalized Deming regression model employing the maximum likelihood estimation (MLE) method \cite{rossi2018mathematical, jensen2007deming}. Assume the errors $\left(e_{x i}, e_{y i}\right)$ and $\left(\varepsilon_{i}, \delta_{i}\right)$ are independent and distributed normally as following,
$$
\varepsilon_{i} \sim N\left(0, \sigma^{2}\right), \delta_{i} \sim N\left(0, \lambda \sigma^{2}\right), e_{x i} \sim N\left(0, \sigma_{x i}^{2}\right), e_{x i} \sim N\left(0, \sigma_{x i}^{2}\right)
$$
where $\sigma_{x i}^{2}$ and $\sigma_{y i}^{2}$ are the known or estimated variances for each observation, $\sigma^{2}$ is the constant variance of error $\varepsilon_{i}$, and $\lambda$ is the pre-determined constant ratio between the variances of two errors $\varepsilon_{i}$ and $\delta_{i}$. Considering the linear relationship between true values $\left(X_{i}, Y_{i}\right)$, i.e., $Y_{i}=\beta_{0}+\beta_{1} X_{i}$, the likelihood function given the transformed values is written as
\begin{equation}
\begin{aligned}
& \mathcal{L}\left(\beta_{0}, \beta_{1}, \sigma^{2}, X_{i} \mid x_{i}, y_{i}, \sigma_{x i}^{2}, \sigma_{y i}^{2}\right) \\
& \quad=\prod_{i=1}^{n}\left(2 \pi\left(\sigma^{2}+\sigma_{x i}^{2}\right)\right)^{-\frac{1}{2}} \exp \left(-\frac{\left(x_{i}-X_{i}\right)^{2}}{2\left(\sigma^{2}+\sigma_{x i}^{2}\right)}\right)\left(2 \pi\left(\lambda \sigma^{2}+\sigma_{y i}^{2}\right)\right)^{-\frac{1}{2}} \exp \left(-\frac{\left(y_{i}-\beta_{0}-\beta_{1} X_{i}\right)^{2}}{2\left(\lambda \sigma^{2}+\sigma_{y i}^{2}\right)}\right).
\end{aligned}
\end{equation}
The log-likelihood function is presented as
\begin{equation}
\begin{array}{r}
\ell\left(\beta_{0}, \beta_{1}, \sigma^{2}, X_{i} \mid x_{i}, y_{i}, \sigma_{x i}^{2}, \sigma_{y i}^{2}\right) \propto 
-\frac{1}{2} \sum_{i=1}^{n}\left[\log \left(\sigma^{2}+\sigma_{x i}^{2}\right)+\log \left(\lambda \sigma^{2}+\sigma_{y i}^{2}\right)+\frac{\left(x_{i}-X_{i}\right)^{2}}{\sigma^{2}+\sigma_{x i}^{2}}+\frac{\left(y_{i}-\beta_{0}-\beta_{1} X_{i}\right)^{2}}{\lambda \sigma^{2}+\sigma_{y i}^{2}}\right].
\end{array}
\end{equation}
The regression parameters are estimated by maximizing the log-likelihood function with an iterative calculation procedure based on the partial derivatives \cite{rossi2018mathematical, peters1978iterative}, for which the derivation is presented in Appendix I. The covariance matrix of regression parameters could be estimated by a bootstrap method \cite{efron1986bootstrap}. 

Overall, this two-stage Deming regression framework is a newly proposed tool to explore the relationship between variables observed with errors of known or estimated variances. Distinguishing itself from conventional regression models, where only the response variable is subject to measurement errors, the proposed framework accommodates the known or estimated variances and other uncertainties of both independent and response variables across diverse scenarios.

\section{Real Application}
\subsection{Motivation and Data source}
The risk of thrombus formation in the left atrium or left atrial appendage is commonly increased for patients diagnosed with atrial fibrillation (AF), which may cause ischemic (embolic) stroke \cite{ivuanescu2021stroke, sharma2015efficacy}. For those patients with high stroke risks, taking oral anticoagulation is a common therapy to help prevent the formation of blood clots. However, the risk of hemorrhage would be increased if taking the anticoagulant and antiplatelet medications, and severe internal or external bleeding may lead to a life-threatening medical emergency \cite{ pengo2001oral, dimarco2005factors}. Consequently, for the same patient, weighing the risks of both stroke and bleeding is essential. To provide recommendations regarding anticoagulation therapy, it's crucial to consider the relationship between these two risks \cite{dimarco2005factors}. By examining the inherent association between bleeding and stroke risks, it becomes possible to assess an individual patient's risks and determine whether their bleeding or stroke risks exceed the expected levels, aiding in medication decisions \cite{goetz2018personalized, sargsyan2022assessment}.

To uncover the underlying relationship between bleeding and stroke risks, we applied the proposed two-stage Deming regression model to a dataset, which encompasses socio-demographic and hospitalization data from 58,088 patients admitted to New Jersey hospitals with AF for the first time between 2000 and 2014. This dataset was sourced from the Myocardial Infarction Data Acquisition System data registry (MIDAS) \cite{wellings2018risk, sargsyan2022assessment, sargsyan2021patient}, a comprehensive statewide database containing hospital discharge data (UB82/UB92) for all patients discharged from non-federal acute care hospitals in New Jersey, along with longitudinal follow-up data spanning up to 30 years.

\subsection{Model fitting}
To apply the proposed two-stage Deming regression model, the initial stage involves estimating the risks of stroke and bleeding individually for each patient, based on the combination of patients' socio-demographic and registry data. The collective outcomes incorporating ischemic stroke and major bleeding were consolidated into a categorical variable, encompassing four levels: stroke only, major bleeding only, both stroke and major bleeding, or neither outcome occurring within one year of the initial $\mathrm{AF}$ admission. The data was modeled by a multinomial logistic regression to estimate the risks. Following stepwise variable selection, eleven risk factors were chosen as predictor variables, which are either dummy variables or categorical variables. These factors include variables such as sex, race (White, Black, other), age ( $<65,65$ to 74, and $\geqslant 75$ years), as well as indicators for diagnoses of heart failure (HF), hypertension (HTN), diabetes mellitus (DM), anemia, chronic obstructive pulmonary disease (COPD), kidney disease (KD), prior stroke, and transient ischemic attack (TIA) \cite{sargsyan2022assessment, sargsyan2021patient}. Given the categorical and dummy nature of the predictors, the probabilities of outcomes for different combinations of risk factors for grouped patients were estimated with standard errors based on the fitted model. The variances of the prediction error terms are calculated by the squared standard errors of estimates. To obtain the estimated risks of bleeding and stroke along with their respective variances or prediction errors, the probabilities of outcomes were aggregated and combined from two categories: stroke or bleeding only, and both stroke and bleeding.

Moving to the second stage, a log transformation was applied to both risks to establish a linear relationship. Given the very small estimated risk values, the incidences per ten thousand were considered by multiplying the risks by ten thousand before log transformation. The variances of errors associated with these log-transformed risks were approximated using the delta method, drawing on the estimated variances of prediction errors obtained in the first stage. To assess the estimated variances from the first stage and determine whether it adequately accounted for overall uncertainties, a criterion represented by Equation \eqref{eq:eq6} was computed, yielding a value of 0.9536. This result indicated that the prediction errors are sufficient to explain the underlying uncertainties of log-transformed estimated values. Based on this evidence, scenario B was selected and fitted using the generalized Deming regression model.

\subsection{Results}
Considering various combinations of dummy or categorical risk factors, 1,749 patient groups were identified, each with estimated stroke and bleeding risks and their prediction variances. These groups with estimated risks served as the input data for fitting a regression line between both stroke and bleeding risks. Within each patient group, the estimated risks, along with their corresponding variances of prediction errors, were assigned different weights, reflecting the frequency of the risk factor combinations in the original dataset which comprised of 58,088 patients. Figure 1 displays the estimates of stroke versus bleeding risks, with (A) showing the log-transformed values and (B) depicting the original incidences per ten thousand. Each data point on the plot represents a specific group of patients with a unique combination of risk factors, and the size of each dot corresponds to its weight or frequency in the original dataset.

To model the relationship between both risks, we employed the proposed two-stage Deming regression model, shown by the black line in Figure 1. For comparison purposes, a weighted least squares (WLS) regression without incorporating the estimated variances of prediction errors from the first stage, was also fitted on the predicted risk values with the data weights \cite{kiers1997weighted}, and represented by the red line in Figure 1. The WLS regression considered only the measurement errors $\delta_{i}$ of the response variable \cite{cornbleet1979incorrect}. Notably, the WLS regression exhibited negative bias, with a smaller absolute value of slope. In the course of fitting scenario B using the generalized Deming regression model with the estimated variances of prediction errors from the first stage, the weights were incorporated into the least-square approach, as indicated by Equation \eqref{eq:eq4}. This was achieved by dividing the prediction error variances by the corresponding weights. Consequently, in Figure 1, points with higher weights were associated with smaller uncertainties in both directions within the proposed two-stage Deming regression model. This improved the performance of the Deming regression line compared to the WLS regression, which did not consider these weight-based uncertainties from the prediction model in the first stage and, as a result, exhibited a biased fit that would underestimate the risk of stroke.

\begin{figure}[h]
\centering
\includegraphics[width=12cm]{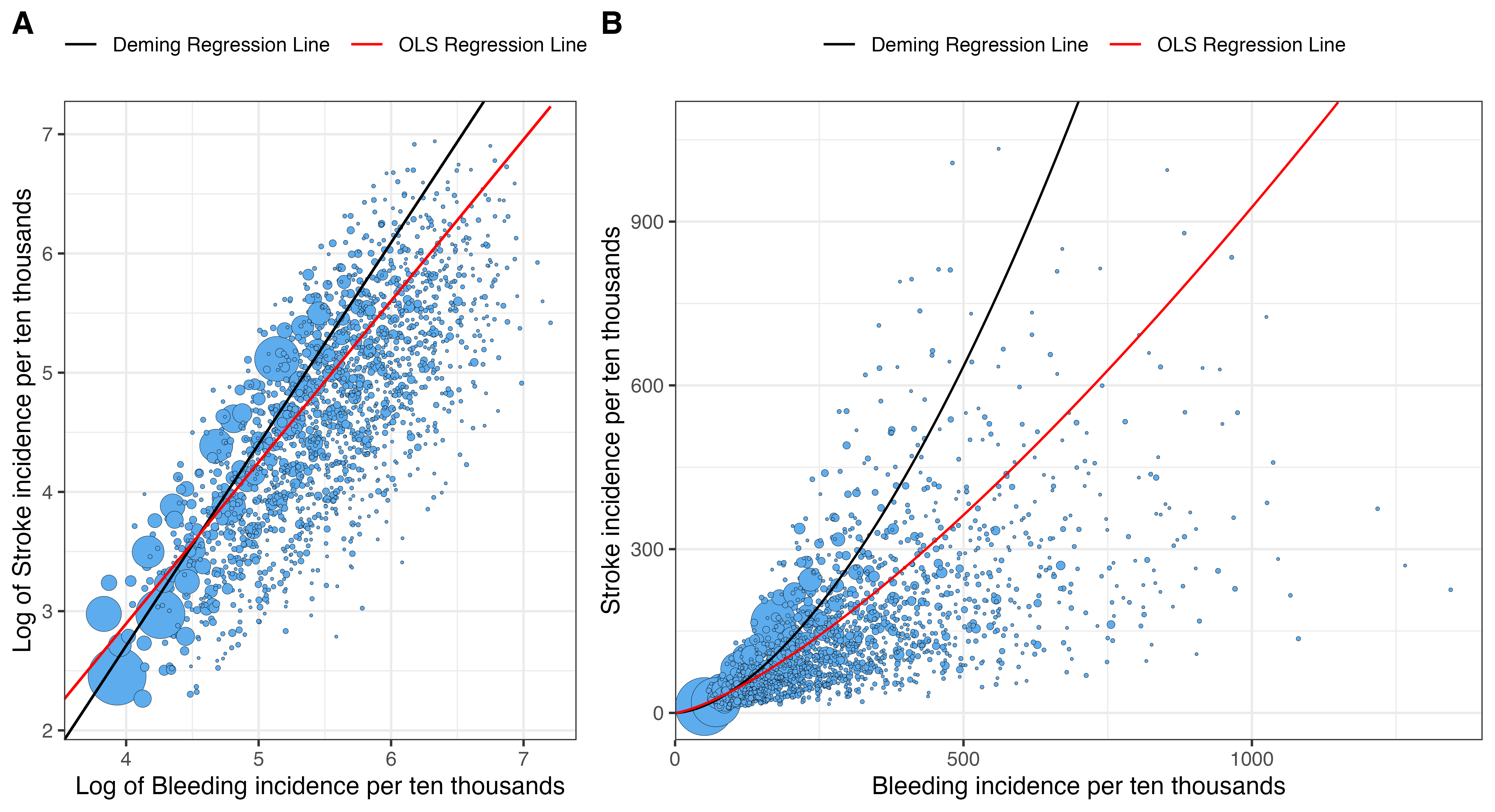}
\caption{\centering{Stroke versus bleeding risks estimates with the fitted Deming regression line (black) and WLS regression line (red) on (B) linear and (A) log-log scales. The size of the dots is proportional to the number of patients with a particular combination of risk factors.}}
\end{figure}

The table presented in Table 1 displays the estimated regression parameters along with their corresponding 95\% confidence intervals (CIs) generated by both the two-stage Deming regression model, and the two-stage framework with the second stage replaced by the WLS regression model. Notably, the WLS regression model exhibited a significant bias, evident in the substantial disparities between the estimates and CIs compared to the two-stage Deming regression model. This bias resulted from the WLS regression's failure to account for prediction error variances on both risks from the first stage. Notice that there is a bias-variance trade-off between two-stage Deming regression and WLS. Overall the bias of WLS is so large that the Deming regression model is preferred. The 95$\%$ CIs from the WLS model are shorter than those from the Deming model because the WLS model does not consider the estimated variances of prediction errors for the risks. 

Furthermore, to assess the effectiveness of the proposed two-stage Deming regression model, two hundred bootstrap samples were created from the original extensive patient dataset. Subsequently, we applied both the two-stage Deming model and the two-stage WLS model to each bootstrap sample to derive $95 \%$ CIs for the regression parameters. The coverages of these two hundred bootstrap CIs over the estimated regression parameters are detailed in Table 1, demonstrating the robust and effective performance of the proposed two-stage Deming regression model. The 95$\%$ CI bootstrap coverages for WLS estimates are surprisingly small because there is a variability of predicted values of risks from the first stage but the variances of prediction errors are not considered in the WLS regression model, while the Deming model takes into account those variances. Moreover, for the 95$\%$ bootstrap CI from the two-stage WLS model, its coverages of the Deming parameter estimates are both $0.05\%$, which means only one coverage out of two hundred CIs.

\begin{table}
\caption{Model Estimates, 95$\%$ confidence intervals (CIs) and 95$\%$ CI bootstrap coverages by two-stage Deming regression model, and by the two-stage framework with the second stage modeled by WLS regression model. }
\centering
\begin{tabularx}{\textwidth}{l|*3{>{\centering\arraybackslash}X}|*3{>{\centering\arraybackslash}X}}
\hline & \multicolumn{3}{c|}{\textbf{Two-stage Deming regression}} & \multicolumn{3}{c}{ Two-stage WLS regression} \\
\hline & Estimated Value & $\mathbf{9 5 \%}$ CI & 95\% bootstrap
CI coverage & Estimated Value & $\mathbf{9 5 \%}$ CI & 95\% bootstrap
CI coverage\\
\cline { 1 - 7 } \textbf{Slope} $\boldsymbol{\beta}_{\mathbf{1}}$ & $1.681$ & $(1.385,1.976)$ & $91 \%$  & $1.355$ & $(1.321,1.389)$  & $25.5 \%$\\
\textbf{Intercept} $\boldsymbol{\beta}_{\mathbf{0}}$ & 
$-4.017$ & $(-5.369,-2.665)$ & $88 \%$ & $-2.532$ & $(-2.699,-2.364)$ & $24.5 \%$ \\
\hline
\end{tabularx}
\end{table}

In addition, prediction intervals (PIs) were constructed to evaluate the model performance \cite{faraw2015practical}. For the regression line generated by the proposed two-stage Deming regression model, prediction intervals were constructed on the mean values given the fitted Deming regression line, and with prediction variances that were computed by considering both the standard errors of regression parameters and the uncertainties associated with both transformed risks. In the context of model fitting under scenario B, the uncertainties linked to both estimated risks could only be explained by the prediction error variances of both risks from the first stage. The derivation of the prediction interval is displayed in Appendix II. Figure 2 displays the 95\% prediction intervals derived from the two-stage Deming regression model. Figure 2(A) shows individual prediction intervals that were calculated based on the prediction error variances of both risks from the first stage for each observation, considering the transformation, i.e., $\sigma^2_{x_i}$, $\sigma^2_{y_i}$. Figure 2(B) employs the square of the mean value of prediction standard errors from the first stage incorporating the transformation, i.e., $\overline{\sqrt{\sigma^2_{x_i}}}^2, \overline{\sqrt{\sigma^2_{y_i}}}^2$ to calculate the prediction intervals. 

To compare the prediction intervals with those from the two-stage framework with the second-stage model replaced by WLS, the prediction intervals were calculated in the same way. However, the estimated parameters with their covariance matrix were replaced by those from the WLS model, with prediction error variances of $x$, i.e., $\sigma^2_{x_i}$, being zero because of the assumption of WLS. The formula is shown in Appendix II. Also, another prediction interval was calculated in a common way for WLS that considers the mean square error (MSE) from the fitted WLS model. Similarly to Figure 2, Figure 3(A) presents the PIs from two-stage WLS using individual prediction error variances of stroke risks from the first stage, and Figure 3(B) displays the PIs using the square of the mean value of prediction standard errors for stroke risk from the first stage based on the observed data. Figure 3(C) shows the PIs considering the MSE from the fitted WLS model. Table 2 provides a summary of the corresponding 95\% PI coverages, considering the weights for both two-stage Deming regression and WLS regression. It's clear that the PIs based on the WLS regression's estimated parameters are more constrained and provide less inclusive coverage when compared to those calculated using the two-stage Deming regression model. Moreover, the $95 \%$ PIs obtained through the proposed two-stage Deming regression model encompassed a significant portion of the actual values, providing strong evidence of the robustness and effectiveness of the proposed methodology. The standard PI of WLS considering MSE has good coverage as expected but the predicted values based on the regression line are biased. 

\begin{figure}[h]
\centering
\includegraphics[width=7cm]{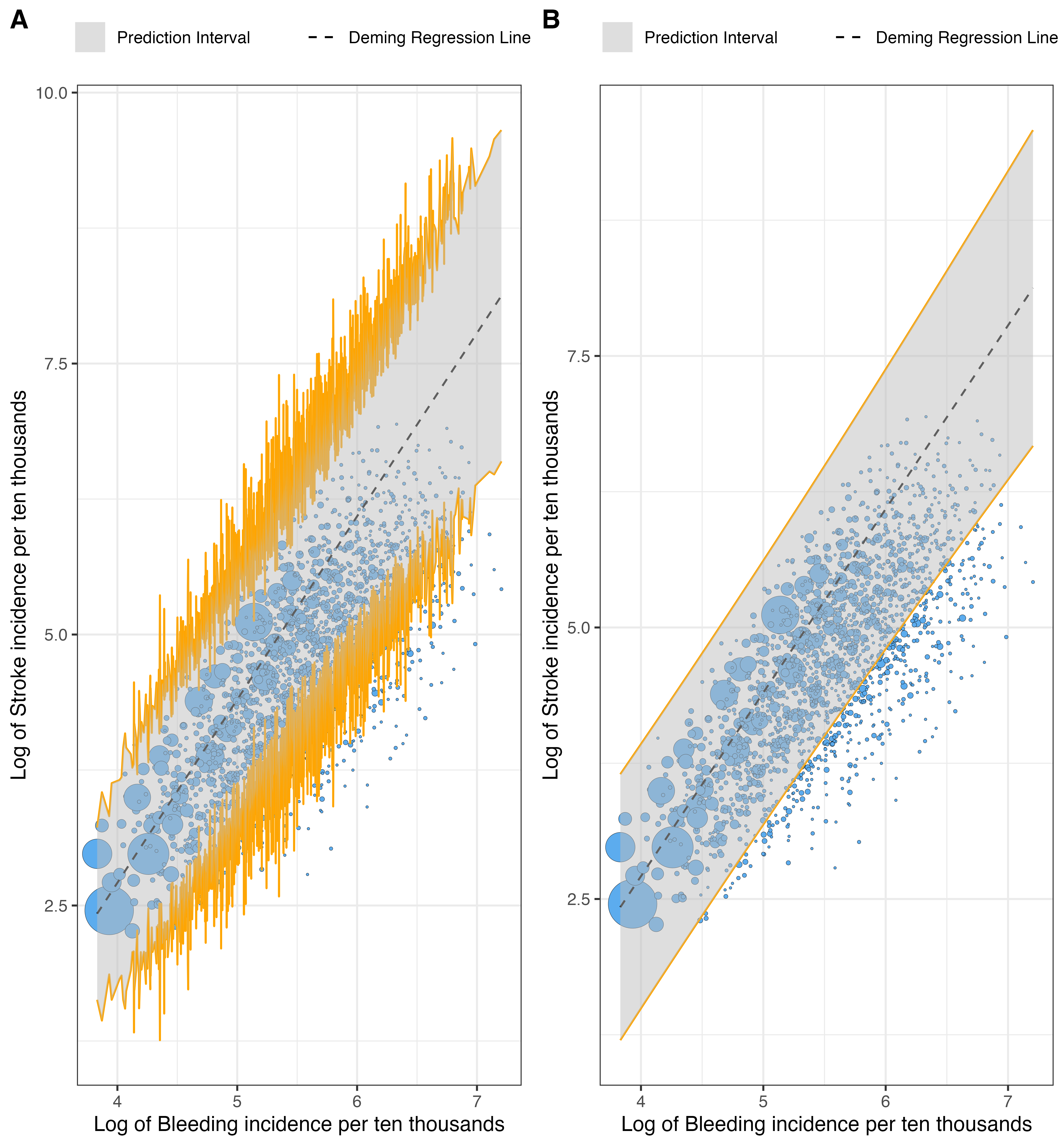}
\caption{95\% prediction intervals by the proposed two-stage Deming model using (A) individual prediction error variances of both risks from the 1st stage considering the transformation, i.e., $\sigma^2_{x_i}$, $\sigma^2_{y_i}$, and (B) the square of the mean value of prediction standard deviations from the first stage incorporating the transformation i.e., $\overline{\sqrt{\sigma^2_{x_i}}}^2, \overline{\sqrt{\sigma^2_{y_i}}}^2$.}
\end{figure}

\begin{figure}[h]
\centering
\includegraphics[width=12cm]{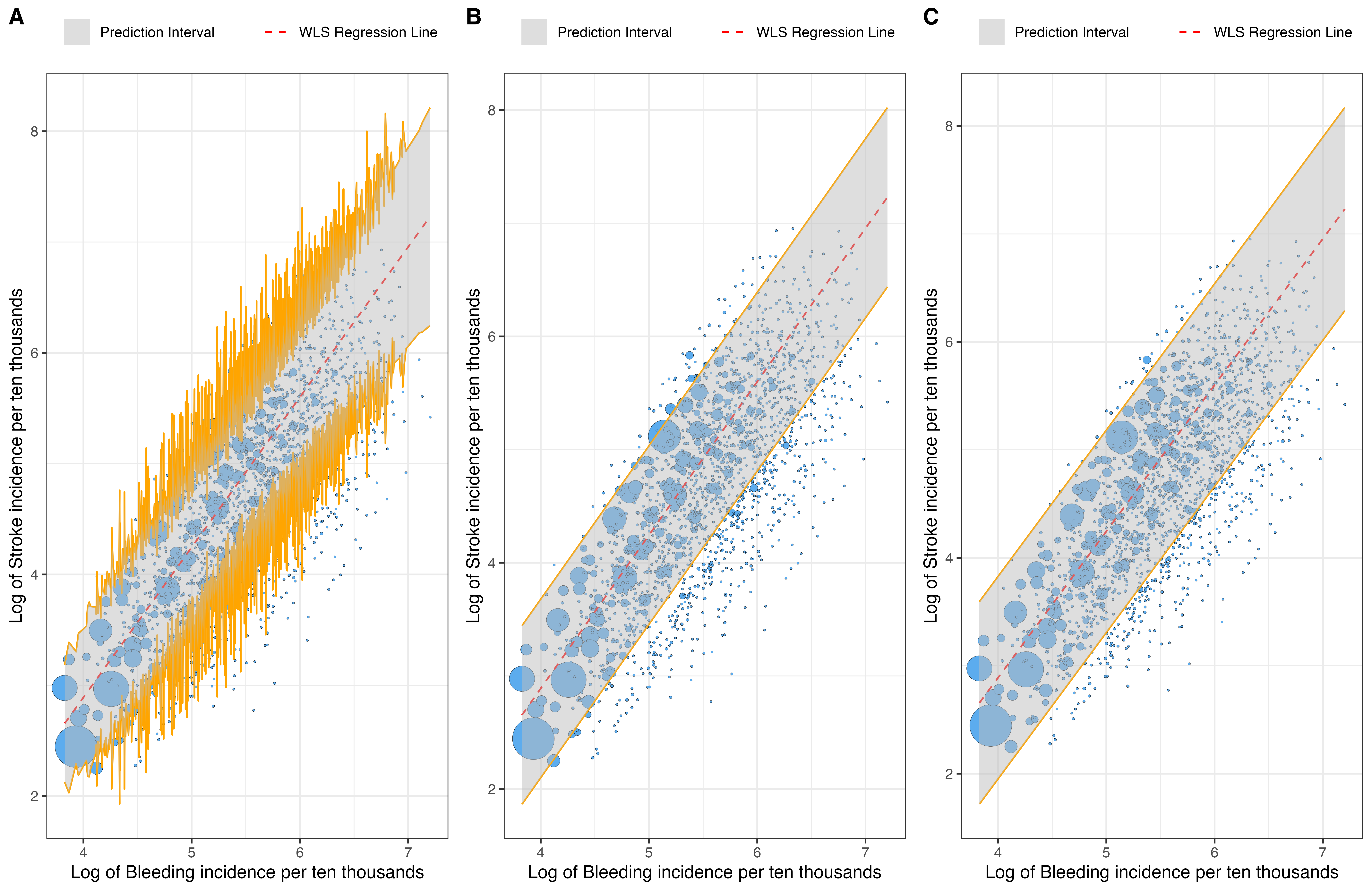}
\caption{95\% Prediction intervals by the two-stage WLS model using (A) individual prediction error variances of stroke risks from the 1st stage considering the transformation, i.e., $\sigma^2_{y_i}$, and (B) the square of the mean value of prediction standard errors of stroke risk from the first stage incorporating the transformation i.e., $\overline{\sqrt{\sigma^2_{y_i}}}^2$, and (C) mean square error (MSE) from the fitted WLS model.}
\end{figure}

\begin{table}
\caption{95\% prediction interval (PI) coverages using individual prediction error variances of risks and the square of the mean value of prediction standard deviations from the first stage, by two-stage Deming regression and WLS regression. The 95\% PI coverage considering mean square error (MSE) by the WLS regression is also shown.}
\centering
\begin{tabular}{l|ccc} 
\hline & \begin{tabular}{c} 
95\% PI coverage \\ using
$\sigma^2_{x_i}$, $\sigma^2_{y_i}$
\end{tabular} & \begin{tabular}{c} 
95\% PI coverage \\ using 
$\overline{\sqrt{\sigma^2_{x_i}}}^2, \overline{\sqrt{\sigma^2_{y_i}}}^2$ 
\end{tabular} & \begin{tabular}{c} 
\end{tabular}\\
\cline { 1 - 4 } \begin{tabular}{c} 
\textbf{Two-stage} \\ \textbf{Deming regression}
\end{tabular} & \textbf{94.892 \%} & \textbf{96.429 \%} & \\
\cline { 1 - 4 }\hline & \begin{tabular}{c} 
95\% PI coverage \\ using $\sigma^2_{y_i}$ 
\end{tabular} & \begin{tabular}{c} 
95\% PI coverage \\ using $\overline{\sqrt{\sigma^2_{y_i}}}^2$ 
\end{tabular} & \begin{tabular}{c} 
95\% PI coverage\\ using MSE
\end{tabular}\\
\cline { 1 - 4 } \begin{tabular}{c} 
\textbf{Two-stage} \\ \textbf{WLS regression}
\end{tabular} & \textbf{71.19\%} & \textbf{93.17 \%} & \textbf{96.67 \%}\\
\hline
\end{tabular}
\end{table}

Utilizing the proposed regression model on a real dataset of patients diagnosed with $\mathrm{AF}$, the underlying connection between the risks of ischemic stroke and major bleeding was accurately identified. Through a comparison of the regression line with individual risk estimates for each patient, it could be evaluated whether the patient had a higher likelihood of experiencing a stroke or bleeding event. This comparison allowed for personalized recommendations regarding the use of anticoagulants to be made, providing valuable guidance to patients. As a result, the insights into the relationship between clinical risks, as determined by the proposed two-stage Deming regression model, serve as a crucial reference for medication decision-making. Similarly, this tool proves valuable in analyzing the association between various clinical risks, particularly when they are weighed against each other, such as disease risks versus the risks of treatment side effects. Leveraging this association analysis of clinical risks enables more personalized and scientifically informed decisions regarding medications and treatments.

\section{Discussion}
In this study, a novel two-stage Deming regression model was introduced and designed to address the regression between variables that are observed with errors of known or estimated variances, a previously overlooked area of research. Within clinical contexts, this model proves to be a valuable tool for conducting association analyses between mutually influential risk factors, such as stroke and major bleeding risks in patients with atrial fibrillation \cite{dimarco2005factors}. By leveraging the insights generated by this new methodology, patients can receive more scientifically informed and personalized recommendations regarding medication usage or treatments, which involves a careful comparison of the underlying risk associations with individual patient circumstances.

Beyond its applications in association analysis of clinical risks, this proposed methodology serves as a versatile tool for regression analyses involving variables observed with errors having known or estimated variances. Within the framework of the two-stage Deming regression model, both independent and response variables were observed or estimated, taking into account their associated error with known or estimated variances in the regression analysis. The inherent uncertainties in both variables were systematically evaluated and discussed across various scenarios, considering the known or estimated error variances. The new methodology allowed for the utilization of both, simple and generalized Deming regression models when dealing with situations where the known or estimated error variances were either disregarded or considered. Additionally, a more intricate version of the generalized Deming regression model was introduced to handle variables that exhibit combinations of known error variances and other unknown sources of variability. While the univariate regression case is primarily discussed in this paper, it is important to note that this approach can be extended to encompass the multivariate case. The paper also introduced a way to calculate prediction intervals for the two-stage Deming regression model using the known or estimated error variances. In addition, an R package is in development to provide a computational tool based on the proposed methodology. 

In the future, there is a possibility of expanding the proposed model to handle regression analyses involving data affected by uncertainties originating from a variety of sources, including both known and unknown factors of variability. For instance, privacy-protected data often contains measurement errors from observed values with unknown variability, as well as privacy errors introduced by a known mechanism \cite{gong2022transparent, evans2023statistically}. Both types of errors require appropriate treatment when exploring relationships within privacy-protected data. Based on the principles of the new methodology, a comprehensive framework could be developed for regression analyses between variables that consider the combination of known and unknown variability. Furthermore, future research directions may involve investigating the covariance between multiple error terms and extending the model to accommodate multivariate scenarios.

In summary, the innovative two-stage Deming regression model proposed in this study could be applied effectively to explore relationships between variables observed with errors of known or estimated variances, while conscientiously considering uncertainties associated with these variables. Its application in the context of association analysis involving clinical risks offers critical guidance to healthcare professionals, enabling them to make individualized decisions about medication and treatments, ultimately leading to improved clinical outcomes. Moreover, the model's adaptability to data with diverse sources of uncertainty, such as privacy-protected data, holds the promise of advancing regression analysis in future research.

\bibliography{ref}

\begin{appendices}
\section*{Appendix I: MLE method for the generalized Deming regression model under Scenario C}
In the proposed two-stage Deming regression model, MLE method was used to fit the generalized Deming regression line under Scenario $\mathrm{C}$ at the second stage. Assuming the normality and independence of the error terms, and considering the linear relationship between true values $\left(X_i, Y_i\right)$, i.e., $Y_i=\beta_0+\beta_1 X_i$, the log-likelihood function of the model was written as
$$
\begin{aligned}
\ell\left(\beta_{0}, \beta_{1}, \sigma^{2}, X_{i} \mid x_{i}, y_{i}, \sigma_{x i}^{2}, \sigma_{y i}^{2}\right) &= -\frac{1}{2} \sum_{i=1}^{n}\left[\log \left(2 \pi\left(\sigma^{2}+\sigma_{x i}^{2}\right)\right)+\frac{\left(x_{i}-X_{i}\right)^{2}}{\sigma^{2}+\sigma_{x i}^{2}}+\log \left(2 \pi\left(\lambda \sigma^{2}+\sigma_{y i}^{2}\right)\right)+\frac{\left(y_{i}-\beta_{0}-\beta_{1} X_{i}\right)^{2}}{\lambda \sigma^{2}+\sigma_{y i}^{2}}\right] \\
&\propto -\frac{1}{2} \sum_{i=1}^{n}\left[\log \left(\sigma^{2}+\sigma_{x i}^{2}\right)+\log \left(\lambda \sigma^{2}+\sigma_{y i}^{2}\right)+\frac{\left(x_{i}-X_{i}\right)^{2}}{\sigma^{2}+\sigma_{x i}^{2}}+\frac{\left(y_{i}-\beta_{0}-\beta_{1} X_{i}\right)^{2}}{\lambda \sigma^{2}+\sigma_{y i}^{2}}\right].
\end{aligned}
$$
where $x_i$ and $y_i$ are the transformed predicted values, $\sigma_{x i}^2$ and $\sigma_{y i}^2$ are estimated variances of prediction errors for each observation, $\sigma^2$ is the constant variance of error $\varepsilon_i$, and $\lambda$ is the pre-determined constant ratio between the variances of two errors $\varepsilon_i$ and $\delta_i$.

The partial derivatives of regression parameters $\beta_0, \beta_1$, assumed constant variances of unknown measurement errors $\sigma^2$, and true (expected) values $X_i$ were derived as
$$
\begin{gathered}
\frac{\partial \ell}{\partial \beta_0}=\sum_{i=1}^n \frac{y_i-\beta_0-\beta_1 X_i}{\lambda \sigma^2+\sigma_{y i}^2} \\
\frac{\partial \ell}{\partial \beta_1}=\sum_{i=1}^n \frac{\left(y_i-\beta_0-\beta_1 X_i\right) X_i}{\lambda \sigma^2+\sigma_{y i}^2} \\
\frac{\partial \ell}{\partial X_i}=\frac{x_i-X_i}{\sigma^2+\sigma_{x i}^2}+\frac{\beta_1\left(y_i-\beta_0-\beta_1 X_i\right)}{\lambda \sigma^2+\sigma_{y i}^2} \\
\frac{\partial \ell}{\partial \sigma^2}=-\frac{1}{2} \sum_{i=1}^n\left[\frac{1}{\sigma^2+\sigma_{x i}^2}+\frac{\lambda}{\lambda \sigma^2+\sigma_{y i}^2}-\left(\frac{x_i-X_i}{\sigma^2+\sigma_{x i}^2}\right)^2-\lambda\left(\frac{y_i-\beta_0-\beta_1 X_i}{\lambda \sigma^2+\sigma_{y i}^2}\right)^2\right]
\end{gathered}
$$
By equating the partial derivatives to zero values and updating estimates step by step, an iterative procedure could be conducted, starting with initial values $\left(\beta_{0,0}, \beta_{1,0}, \sigma_0{ }^2, X_{i, 0}\right)$, to seek estimates that maximize the log-likelihood function. The bootstrap method can be utilized to estimate the covariance matrix of the regression parameters.

\section*{Appendix II: Prediction Intervals for Two-stage Deming regression model and WLS model}

Under the proposed two-stage Deming regression model, based on a new observation $x_{new}$ with known or estimated variances of the errors $e_x$ and $e_y$ from the first stage, the predicted $y$ denoted by $y_{new}$ is given by

$$y_{new} = \hat{\beta}_0 + \hat{\beta}_1 x_{new} + e_y = \hat{\mathbf{\beta}}X_{new} +e_y,$$
where $X_{new} = \begin{pmatrix}
    1\\
    x_{new}
  \end{pmatrix}$.

Given the independency between $X_{new}$ and $e_y$, as well as the independency between $X_{new}$ and $\hat{\beta}$, the prediction variance of $y_{new}$ was derived as following, 
$$
\begin{aligned}
Var(y_{new}) & = Var(\hat{\mathbf{\beta}}X_{new}) + Var(e_y) \\ 
& = X_{new}^T Cov(\hat{\mathbf{\beta}})X_{new} + \hat{\mathbf{\beta}}^T Cov(X_{new})\hat{\mathbf{\beta}}+ \text{Trace}(Cov(\hat{\mathbf{\beta}}) Cov(X_{new})) + Var(e_y)\\
   & = X_{new}^T Cov(\hat{\mathbf{\beta}})X_{new} + \hat{\beta}_1^2 Var(x_{new}) + Var(\hat{\beta}_1)Var(x_{new}) + Var(e_y) \\
    & = X_{new}^T Cov(\hat{\mathbf{\beta}})X_{new} + 
    (\hat{\beta}_1^2 +Var(\hat{\beta}_1))Var(x_{new}) + Var(e_y) 
\end{aligned}   
$$
The covariance matrix of the estimated regression parameters can be obtained from the Deming model. The variance of $x_{new}$ is known from the first stage, and the variance of the error term $e_y$ can be estimated by the square of the mean or median value of the known standard deviations of $y$ from the first stage based on the observed data, or determined by prior knowledge. 

Therefore, the  $(1-\alpha)100\%$ prediction interval for $y_{new}$ based on the two-stage Deming regression model is written as
$$\hat{y}_{new} \pm t_{(1-\alpha/2, n-2)}\sqrt{Var(y_{new})},$$
where $\hat{y}_{new} = \mathbb{E}(y_{new}) =  \hat{\beta}_0 + \hat{\beta}_1 x_{new}$. 

To compare the prediction intervals with those from the two-stage framework with the second-stage model replaced by WLS, the prediction intervals were calculated in the same way, with the estimated parameters with their covariance matrix replaced by those from WLS model, and with $Var(x_{new})$ being zero because of the assumption of WLS. Moreover, another prediction interval was also calculated in a common way for WLS that considers the mean square error (MSE) from the fitted WLS model. The results of prediction interval coverages from both methods were shown in Table 2, Figure 2, and Figure 3.

\end{appendices}

\end{document}